# The Technology Transfer of the Italian Public Research System: the Case of the National Research Council of Italy[1]

*Giovanni Abramo*

Laboratory for Studies of Research and Technology Transfer
Department of Management, School of Engineering
University of Rome "Tor Vergata"

and

Italian Research Council

## 1. Introduction

The more rapid growth of some industrialized nations is, in part, the result of an effective complementary role for government in providing the domestic economy with an adequate technology base as well as broad and rapid utilization of this base (Tassey, 1991).

While European Union (EU) scientific and technological institutions show levels of research productivity comparable with North America's ones, their contribution to the competitiveness of the European industry lags quite behind. The reason lies in the EU weak integration between public research institutions (PRIs[i]) and industry (Ben-David, 1977), as compared to North America's (Geiger, 1988; Rosenberg and Nelson, 1994; Stokes, 1997). Only at the end of the 80s, induced by a strong anxiety about EU competitiveness in science-based industries, EU and, at different levels, EU single member states have emphasized the importance of academic technology transfer (Lowe, 1993, Peterson e Sharp, 1998). Not only have the United States and Canada become aware of the importance of the integration of the two systems long before the EU, but they have acted accordingly with an impressive rapid pace. In 1980 approximately 20 US universities only had a technology licensing office (TLO); by 1990 the number was 200, and by 2000 nearly every major research university had one (Colyvas et al., 2002). The *Licensing Survey: Fiscal Year 2003* of 2004, by the American Association of University Technology Managers, shows that since the passage of the Bayh-Dole Act in 1980, the number of patents filed yearly by North American PRIs has increased more than tenfold and, at the same time, there was also an increase in the number of patents transferred to industry. Before the Bayh-Dole Act, 5% only of US PRIs patents were licensed to industry (Mowery and Sampat, 2005). Currently, about 60% of US PRIs patents is licensed out or optioned, of which 65% to small companies (with fewer than 500 employees) or spin-offs. 374 spin-offs were created, 79% of which in the region where the underlying technology was developed. At the same time, income from licensing has grown almost tenfold since 1991.

This very rapid patenting, licensing and revenue growth rates, has led few scholars to investigate whether the surge of commercialization activities has fundamentally changed the nature of US academia for the worse. One of the first concerns addressed was whether an emphasis on the commercialization of knowledge assets might entail the risk of tipping the research balance from lower to higher patent-intensive sectors and from fundamental to applied research. This prospect, however, was denied by empirical evidence by Mowery et al., 2001 and Thursby and Thursby, 2002. Nelson, 2004, observed that uncontrolled patenting and exclusive licensing might be counterproductive for research itself: knowledge usually follows gradual developmental patterns, and patenting intermediate findings or tools useful in following stages of the research process can slow down the advancement of knowledge and technologies, especially in the area of fundamental research. Washburn, 2005, analyzed a number of cases of negative externalities associated with indiscriminate patenting and opportunistic licensing behavior by US universities. While this US introspection is underway many industrialized countries are changing their university and legal systems and making substantial financial investments to emulate the environment that Bayh-Dole seems has created. Actually, Mowery and Sampat, 2005, recently stated that much of the growth in licensing and university "spinoffs" that has occurred since the passage of the Bayh-Dole Act almost certainly would have occurred in the absence of this piece of legislation, because of the long-standing and relatively close relationship between US universities and industrial innovation. They go further to question whether the emulation of the Bayh-Dole Act, occurring in a number of Oecd (Organization for economic co-operation and development) countries, is sufficient or even necessary to stimulate higher levels of technology transfer. Instead, reforms to enhance inter-institutional competition and autonomy within national university systems appear to be more important.

While a higher and higher number of EU countries continue to strive to emulate US success in harnessing the intellectual output of its academic institutions, Italy still seems to lag behind, in terms of both relevant actions and performances

A recent report on firms' innovativeness by the Italian Statistics Institute, Istat (2003), shows that Italian companies which had introduced product or process innovations in the period 1998-2000, rank PRIs last among ten different sources of information for the innovations introduced. As for the technology transfer performances of Italian PRIs, very little can be found in the literature. The very few empirical analyses tend to concentrate on scientific output (publications) of Italian PRIs research activity (Bonaccorsi and Daraio 2002, Orsenigo 2002). However, indicators of scientific production, such as scientific papers, citations, impact factor, etc. if on the one hand may result useful towards the appraisal of the quality and efficiency of research systems, on the other have little significance with regard to the evaluation of the impact of public research on domestic industry's competitiveness. While scientific papers play a significant role as a source of information for companies' innovative activities they hold little potential as sources of firms' competitive advantage, because of the "public" essence of this way of encoding knowledge. Studies on Italian PRIs patent outputs, are rather limited. Important exceptions are Piccaluga and Patrono (2001) and Balconi et al. (2003) that analyze Italian PRIs patents filed at the European (EPO) and American (USPTO) patent offices. Piccaluga et al. (2002), extend the empirical analysis by comparing with France and Spain. Baldini et al. (2003) delve into the effects of institutional changes on the patenting activities of Italian universities. Abramo (1998) was the first one to engage in the analysis of licensing performance, although limited to the Italian Research Council (Cnr), the main public research institution in the country. Recently, Abramo and Pugini (2005) extended the licensing investigation to the academic system. The paucity of empirical studies reveals, on one side, the little attention to the problem at different levels, political and institutional first of all; and on the other, a great difficulty in acquiring the relevant data for appropriate elaboration.

This paper will analyse the patenting and licensing activities of the Italian Research Council from 1996 to 2001[ii]. A benchmarking of Cnr main performance indicators is then conducted vis-à-vis the Massachusetts Institute of Technology (Mit) for the period 1999-2001. Finally, through a time series analysis based on a previous study from the same author (Abramo, 1998), a drop in patenting performance will be pointed out, followed by the identification of its main determinants. The purpose of the study is twofold: i) to provide a complete picture of how commercialization activities of Italian PRIs stand vis-à-vis the best practices; and ii) to identify the likely determinants of low performances, which may provide practical implications for Italian policy makers with respect to how to foster technology transfer.

The articulation of the article is as follows. Section 2 will deal with how PRIs can contribute to industry competitiveness through technology transfer. Section 3 will analyze Cnr technology transfer. Section 4 will contrast it with Mit. Section 5 will present a Cnr performance time series analysis. Finally, section 6 will provide a discussion of the main determinants of the findings.

## 2. Assessing the capability of the public research system to sustain domestic industry's competitiveness.

The generic purpose of PRIs is to sustain national competitiveness and socio-economic development, through the advancement of science and technology progress. The appropriate indicators and their relative importance to assess the pursuing of this purpose descend directly from the development policy of the government (in general, customer benefits, net production investments and new jobs[iii] induced by the utilization of public R&D outputs by industry, but also specific sectorial or regional development, etc.).

It is possible to disaggregate the process by which public R&D leads to socio-economic development into its three major stages: R&D by PRIs, technology transfer from PRIs to industry and technology exploitation by industry. Then, the efficiency of the public technology infrastructure, or the social return to public R&D investment, is timely measurable in terms of: i) R&D productivity, i.e. R&D output divided by R&D investment; ii) technology transfer productivity, i.e. R&D output actually transferred to the production system divided by total R&D output (speed of technology transfer and amplitude of diffusion are other indicators of technology transfer efficiency); and iii) transferred technology yield, i.e. social benefits as defined above divided by transferred R&D outputs.

The problem is to identify appropriate indicators of R&D output which have direct linkages to industry competitiveness. We know that competitive advantage at business level is based on the so called "distinctive competences", that is firm's strategic resources which, in addition to being superior to those of competitors', are also difficult to imitate, transfer and substitute. As a consequence, in order for the public R&D output to be able to induce distinctive competences within domestic companies, it is not enough that knowledge advances be superior to those produced by other national research infrastructures, but also that: i) the mode in which new knowledge is encoded be such to be difficult to imitate or transfer without the owner's will, and ii) the properly encoded new knowledge be actually transferred to industry.

The mode of encoding new knowledge most apt to create competitive advantage is then intellectual property protection, i.e. patent and the like. Actually, tacit knowledge and trade secrets, which may be transferred through consulting or training, are other important sources of competitive advantage as well. Scientific papers, because of their public nature, are not a source of competitive advantage. Furthermore, because it has been shown that large companies recur to scientific literature as a source of information much more than small companies (Cohen et al., 2002), those countries whose industry structure is characterized mainly by large companies (which is not the case of Italy) benefit more than others from scientific papers.

The simple formula that I apply to measure the effectiveness of a public research system in sustaining domestic industry competitiveness and socio-economic development is then:

$$\text{Return to R\&D investment} = \frac{P}{\text{Inv}_{R\&D}} * \frac{L}{P} * \frac{SB}{L} = \frac{SB}{\text{Inv}_{R\&D}}$$

Where P represents the number of patents and the like generated by the R&D investment, $\text{Inv}_{R\&D}$; L is the number of patents licensed out; and SB represents the social benefits (customer benefits, net production investments and new jobs, etc.) induced by the exploitation of the licenses by industry.

Examining returns in terms of the above three factors, namely R&D productivity, technology transfer productivity and socio-economic yield, has organizational advantages too. R&D productivity is determined largely by the efficiency of the R&D labs. Similarly, technology transfer productivity is determined largely by the efficiency of the technology transfer office (i.e. the marketing and sale function of the organization). Socio-economic yield, on the other hand, in addition to the innovation potential, is largely dependant on industry structure. Actually, the measurement of the first two factors, whose product is named "quantitative return to R&D investment", may be sufficient to appraise the relative efficiency of public research systems in sustaining domestic industry competitiveness.

This methodology and the relevant formula are by no means exhaustive and not intended to pursue an accurate quantitative measure of socio-economic returns to public R&D investments. In fact, only patent and patent license are considered as indicators of, respectively, R&D output and technology transfer. Moreover, the yield factor is not easy to measure because of the intrinsic difficulty of identifying all innovation externalities and quantifying social benefits. Although patent-based indicators are probably the most frequently used measure of technology output, one should bear in mind a few drawbacks: first, institutional differences, distinct appropriability strategies, and different propensity to patent across sectors may bias the international comparisons. Second, the value distribution of patents is very skewed, with many patents having no industrial application and few patents very high returns. Third, changes in patent laws make it difficult to analyze long run time trends. Last but not least, patent indicators are constructed on the basis of applications filed at national patent offices, having different examination procedures and an "home advantage" bias.

Nevertheless, this methodology can be a powerful managerial tool to benchmark national technology systems, as the following analysis hopefully will show.

## 3. The case of the National Research Council of Italy

With an R&D expenditure as a percentage of GDP of 1,16%, Italy ranks 22$^{nd}$ among the 30 Oecd countries. However, because of its comparatively high GDP, Italy ranks 8$^{th}$ in the world in terms of gross domestic expenditure on R&D. The comparison of R&D expenditure by source of funds and sector of execution shows that Italy has the highest percentages among G7 countries: 50.8% of R&D is funded by the public sector and 51% is executed at PRIs. The average figure of G7 is around 30%. Italian research scientists result very productive but, much more than their foreign colleagues, they tend to encode new knowledge in the form of publications rather than patents. From 1995 to 2000, Italy scored a record annual publication growth rate among G7 countries, which has now resulted in the second (after UK) highest number of publications per researcher (Oecd, 2004). Conversely, patent applications by universities remain relatively scarce, as the whole of Italian universities file as few patents as the university of Wisconsin alone. Even if all other government research laboratories are counted in, the total number of PRIs filed patents stays below the Mit alone. The gap in patent intensity, then, is alarming: in 2002, patent intensity of Italian universities amounted to merely 4 patents per 1,000 researchers, as against 22 in the UK, and over 40 in the US as early as in 1999 (Abramo and Pugini, 2005).

To complete the picture and try to better understand the causes of the scant entrepreneurship attitude of Italian PRIs an analysis of the patenting and licensing activities of the Italian Research Council has been carried out.

What is Cnr? It is the main public research institution in Italy in terms of size, budget, geographical coverage and research spectrum. Cnr promotes, co-ordinates and carries out research in almost all fields of science, through its over 100 research institutes[iv], with a staff of approximately 8,000 of whom 4,000 are research scientists. In addition, in order to favor the transfer of its research outputs in 1980 a Cnr technology licensing office was set up.

Why is Cnr an appropriate benchmark to represent Italian PRIs in international comparisons? Not only is Cnr the largest public research institution in the country (12% of total R%D expenditure), above all it shows the largest patent portfolio and patent intensity among PRIs in the period 1982-2001, owning 59% of all EPO and USPTO patents filed by the whole Italian public research system, i.e. 9 times as many as Italian universities altogether[v] (Piccaluga and Patrono, 2001). Comparing then the largest and most productive public research institution in the country with a leading American research university, Mit, will provide a good perspective of where the Italian public research system stands vis-à-vis one of the world leaders.

*3.1. Cnr productivity analysis*

In the period 1996-2001, Cnr average yearly research expenditure was € 747.51 million at constant prices 2001 (table 1). 76% of the total expenditure, i.e. € 566.55 million was allocated to institutional research activities carried out at Cnr Institutes; 13%, i.e. € 94,05 million, was allocated to the so called "Targeted Projects"[vi] (TPs); and the remaining 12%, i.e. € 86.92 million, to general expenses including buildings, training, and technology transfer. In the same period, Cnr filed 257 patents[vii], i.e. 43 patents per year on average. 21% of all patents stemmed from TPs. 67% of all patents was filed in Italy; 26% was PCT[viii], 4% was USPTO only, 2% was EPO only, 4% was others. Cnr was the exclusive owner of 77% of patents; 15% was jointly owned by Cnr and other PRIs; while 8% was jointly owned by Cnr and private companies.

R&D productivity in the period at stake was 5.71%, that is € 100 million R&D investment produced on average 5.71 patents. In particular, productivity within Targeted Projects was 8.62%, therefore above average, as could be expected: in fact TPs are more oriented towards applied and industrial research. Productivity within institutional research was 5.25%.

Table 2 shows the number of patents licensed every year to industry. The total number of patents licensed in the period under exam amounted to 50, which corresponds to a yearly average of 8.33. 6 out of them, i.e. 12% of all patents licensed out, were licensed abroad. Technology transfer productivity, i.e. the ratio of patents licensed out to patents filed in the period, was 19.46% (table 3). 16% of all patents licensed stemmed from TPs; while the remaining 84% from institutional research. Technology transfer productivity of TP patents was 14.55%, below average; while technology transfer productivity of patents arising from institutional research was 20.79%. This evidence is quite surprising.

Out of the 50 patents licensed in the period under exam, 42% was filed in Italy, 38% was PCT, 6% was EPO, and the remaining 14% was filed elsewhere. Technology transfer productivity of patents filed in Italy only was below average, 12.28%; while technology transfer productivity of PCT patents was 28.36%. This can be explained with the Cnr policy of extending patent protection abroad only if a potential licensee has formally shown interest in the extension.

As for the ownership, 78% of patents licensed belonged exclusively to Cnr; 10% was jointly owned by Cnr and other PRIs; and 12% was jointly owned by Cnr and industry. Technology transfer productivity of Cnr wholly owned patents was 19.70%; of Cnr-PRIs jointly owned patents was 12.82%; of Cnr-industry owned patents was 30%. Although this last finding is above average, it is quite surprising that 70% of patents filed by private companies jointly with Cnr were not exploited by the relevant companies.

A different measurement of technology transfer productivity may be obtained by the ratio of licensed patents to patent portfolio (i.e. licensable patents). In fact, every year it is possible to license not only inventions patented in that year, but also all previous years' patents which have not been abandoned or exclusively licensed. Every year, on average, 1.5% of Cnr patent portfolio is licensed out. The technology transfer productivity referred to patent portfolio depends, among other things, on how wisely the patent portfolio has been built up. On the one hand, a large portfolio enhances licensing-out probability; on the other hand, patents' renewal fees increase. The latter increases proportionally with the age of patents: the older the patent, the higher the yearly renewal fee. Furthermore, because of technological obsolescence, the licensing-out likelihood is a diminishing function of the age of patents.

Two other important indicators describe technology transfer effectiveness with regard to socio-economic development, namely diffusion of innovation and rate of technology transfer. The analysis of the distribution of non-exclusive licenses and exclusive licenses shows that out of 50 patents licensed out, 38 (76%) were granted as exclusive licenses. The remaining 12 patents gave rise to 40 non-exclusive licenses. Out of 78 overall licenses, 21 (27%) were granted abroad. As for the rate of technology transfer, the shorter the lag between patent filing and licensing, the higher the impact on competitiveness and socio-economic returns. In the period considered, 45% of licenses were granted within one year from filing; 84.2% within three years; and only 15.8% of licenses were granted after three years of filing.

At this point it is possible to calculate the overall Cnr quantitative return to R&D investment, as the ratio of licensed patents to R&D expenditure, or the product of research productivity by technology transfer productivity. In the period 1996-2001 the average Cnr quantitative return to R&D investment was 1.12%, which means that € 100 million R&D investment led to 1.12 inventions transferred to industry.

### 4. Cross-section analysis

The previous section showed an apparent inertia of technology transfer from Cnr to industry. The question is whether this inertia is intrinsic or typical of the Italian system. To try to answer this question it has been carried out a benchmarking exercise between Cnr, i.e. the most productive (in terms of patents) PRI in Italy, and the Massachusetts Institute of Technology (Mit), one of the top research universities in the US.

Mit is an appropriate benchmark as it shows characteristics, pertinent to the subject of this study, similar to Cnr's. Like Cnr, Mit carries out research in a number of disciplinary sectors and at different levels: basic, applied and technological research. It is similar in size: It employs around 7,800 employees, of which 1,500 are professors/lecturers and 1,700 are researchers. Also, it has a yearly research budget very close to Cnr's. Furthermore, like Cnr, Mit devotes a similar share of its R&D budgets to humanities and social research (sectors where research output is hardly encoded as patent). Finally, it has an internal Technology Licensing Office (TLO). The comparative analysis has been carried out for the period from 1999 to 2001. The relevant findings are shown in table 4.

The Mit yearly average R&D budget in the period was € 735 million (at 2001 constant prices)[ix], while Cnr's was € 763 million. On average 151 Mit patents[x] were issued yearly, and the Mit TLO licensed on average 89 patents per year[xi] out of its overall patent portfolio. Mit average R&D productivity then, was 20.54%; while technology transfer productivity was 58,94%. In 2001, the average R&D productivity of North American PRIs was 19,25%, while technology transfer productivity[xii] was 59,57%. Cnr, instead, patented in the same period an average of 41 inventions[xiii] per year and licensed on average 6.33 patents per year out of its patent portfolio. Cnr R&D research productivity was then 5.33%, while technology transfer productivity was 15.57%. Licenses considered here did not include trademarks and copyrights[xiv].

In comparative terms then, Cnr patent output was 27% of Mit's; while Cnr licensed out patents were only 7% of Mit's. It is worth noting that up to 1985, Cnr patent output was higher than Mit. Comparing both institutions' productivity indexes, it results that Cnr R&D productivity was around 26% of Mit which means that, R&D expenditures being equal, Cnr produced 26% of Mit patents. Cnr technology transfer productivity was also around 26% of Mit's that is, the number of licensable patents being equal, the Cnr TLO was able to license out about 26% of patents which was able to license out the Mit TLO.

By multiplying the above two factors it is possible to have a measurement of the quantitative return to R&D investment (number of licensed patents divided by R&D expenditures) and, therefore, an indication of the relative efficiency of the two research organizations. Mit showed an average return of 12.11%, which means that € 100 million R&D investment pours out 12.11 inventions into the production system. Instead, Cnr showed a return of 0.83%, which means that to pour out 8.3 inventions it takes € 1 billion. In relative terms, R&D expenditures being equal, Cnr poured out only 6.85% of the inventions which Mit was able to generate and transfer.

With reference to a previous benchmark analysis of Abramo (1998), referred to year 1994, the current findings show that Cnr R&D productivity as compared to Mit has declined to less than half; while relative Cnr technology transfer productivity has doubled. The overall relative Cnr quantitative return to R&D investments has declined by 1% with respect to 1994.

Finally, as far as the socio-economic return to R&D investment is concerned, a study carried out in early 1995 by Pressman et al. estimated that Mit active, exclusive, patent license agreements had induced just under US$ 1 billion investments by the commercial sector toward the development and early commercialization of licensed inventions, and over 2,000 jobs had been created and/or sustained as a direct result of these licenses[xv]. This estimate was conservative as many licensees had not yet introduced the relevant products into the market: then, further development, production and marketing investments would be required.

Assuming that analogous amounts of investment would be required on the Italian commercial sector toward the development and early commercialization of Cnr-licensed inventions, it is easy to have an order of magnitude of the amounts of investments and new jobs forgone because of Cnr lower technology R&D and technology transfer productivities (6.85% of Mit).

At financial level, the Mit TLO showed average yearly revenues of € 46.61 million, and expenditures on patents of € 6.58 million. The Cnr TLO showed average yearly revenues of € 640,000 and expenditures on patents of € 390,000. Cnr TLO revenues represented 1.37% of Mit; while Cnr TLO profit margin was 0.62% of Mit. Mit TLO revenues showed a constant growth, representing in 2001 11% of Mit total R&D expenditures. A growing share of Mit TLO revenues derived from sale of Mit spin-off stakes. In the period 1999-2001 Mit created on average 23 spin offs yearly, while Cnr created its first 3 spin offs only in 2002.

5. **Time-series analysis**

In order to evaluate the evolution of Cnr technology transfer performance a time-series analysis of the main productivity indicators has been carried out. Findings of this investigation have been compared to those of a previous study of Abramo (1998) referred to the period 1989-1995. The time-series analysis is shown on table 5.

The period 1996-2001 shows a contraction (-9%) of R&D expenditure with respect to the period 1989-1995, which confirms a trend: the period 1989-1995 witnessed itself a decrease (–7%) of R&D expenditure with respect to the period 1981-1988. The contraction of R&D expenditure hit above all Targeted Projects (-19%), and to a less extent Institutional Research (-8%).

Patent output decreased by 41%, passing from a yearly average of 73 patents in the period 1989-1995 to 43 in the later period. The major contraction regards patents stemming out of Targeted Projects (-74%), while the number of patents stemming out of Institutional R&D diminished by 13%. As to be expected, the number of patents jointly owned by Cnr and private companies passed from a yearly average of 8.5 in the period 1989-1995 to 3.3 in the period 1996-2001 (-61%).

Since in the latter period patent output decreased more than R&D expenditure, R&D productivity worsened (-34%), passing from 8.85% in the former period to 5.85% in the latter one. In particular, Targeted Projects R&D productivity decreased by 67%, while Institutional Research productivity decreased by 7%. Although Targeted Project R&D productivity was 1.6 times as high as Institutional R&D productivity, the proportion has considerably diminished with respect to the period 1981-1995, when it was 4.6 times as high. It is worth underlying that such decrease is not due to a growth of Institutional R&D productivity, which, as already said, diminished by 7%.

While patent output and R&D productivity diminished in the latter period, the number of patents licensed out slightly grew (+2%). As a consequence, technology transfer productivity considerably rose (+74%). In particular, technology transfer productivity concerning patents stemming out of Targeted Projects rose by 81%, while technology transfer productivity referred to Institutional R&D rose by 19%. Technology transfer productivity measured as the ratio of patents licensed out divided by the whole patent portfolio rose only by 11%. This is due to the fact the patent portfolio diminished (-7%) comparatively less than patents filed (-41%) in the later period.

The overall quantitative return on R&D investment rose by 13%, passing from 0.99 patents licensed out per € 100 million R&D investment in the former period, to 1.12 in the latter period.

Few findings of the time-series analysis deserve to be delved into. First of all, the considerable drop of patent output in the latter period. This phenomenon can be the effect of essentially four causes: i) the drop of R&D expenditure; ii) the drop of the share of R&D funds devoted to Targeted Projects, which show the higher research productivity; iii) the shift of Targeted Projects to less "patent fertile" research fields; and iv) the new Cnr patent policy of 1995[xvi], according to which the burden of patent expenses had been decentralized from corporate level to research Institutes and Targeted Projects, causing a competition within research centers in the allocation of scarcer and scarcer funds between research and patenting expenses. The drop of R&D expenditure (-9%) can explain only a part of the larger drop of patent output (-41%). Assuming constant return to scale along the relevant patent-R&D expenditure arch, the input drop could cause a patent output decrease of 9% at most. It must be noticed that to an analogous input drop in the period 1989-1995 (-7%), it corresponded a raise of patent output of 4%. The impact of the lower share of funds devoted to Targeted Projects, which passed from 15.8% of total Cnr R&D expenditure to 14.2%, is evidently negligible. The main determinants of the output drop remain then the last two factors, whose relative weight is difficult to quantify. A comparative analysis of the areas of investigation of TPs showed, in fact, a shift in the latter period from the Pasteur and Edison quadrants to the Bohr quadrant (Stokes, 1997), i.e. from research that, be it basic or applied, had commercial application to areas less fertile in terms of patent output. The decentralization of resource allocation decisions from Cnr headquarters to research institutes, i.e. research scientists boards, inevitably penalized patenting investments vis-à-vis research expenditures, given the simultaneously scant patenting culture of Italian scientists and the growing hunger for research resources.

Finally, although the number of patents filed has dropped, the number of patents transferred has slightly risen. This may be the effect of the "Darwin law" induced by the new above said patent policy, which led to a stricter selection process and a better quality of patented inventions, and/or an improved efficiency of the Cnr TLO. Both causes are plausible, although it must be noticed that 40% of patents licensed out in the period 1996-2001 were filed before the introduction of the new patent policy, and that the technology transfer productivity referred to patent portfolio rose only by 11%. There may be also a diminishing return to scale effect (to an increase in the number licensable patents it corresponds an increase in the number of licenses less than proportional), as shown by Thursby and Thursby (2002) in the American case.

## 6. Discussion and conclusion

The foregoing analysis, far from being aimed at a quantitatively precise assessment of the technology transfer performance of that significant component of the Italian public technology infrastructure represented by Cnr, allows to appreciate the dimension of the gap with Mit and the US university system in general. A gap which extends even more remarkably to the whole Italian public research system, if one considers that Cnr technology transfer productivity is 46% higher than overall Italian universities (Abramo and Pugini, 2005). This gap can be considered neither negligible nor unnoticeable, notwithstanding assumptions, approximations and limitations that comparisons of this kind intrinsically embed.

The gap shows in both patenting and licensing dimensions of technology transfer. Furthermore, along the former dimension the Cnr time series analysis shows a worrisome drop in performance. How do we explain then such low patenting intensity? Is it just a matter of the higher quality of American research scientists as compared to Italians? This may indeed explain part of it, but if this holds entirely true how would we explain then the higher Italian research productivity in terms of publications per researcher? And when would that difference in quality have occurred, if till 1985 the number of Cnr patents was higher than Mit? Nor can the patenting gap be explained with different patent legislations in the two countries: they were substantially the same in the period under exam. The major determinants of such a huge gap are to be found somewhere else, in particular among cultural, strategic and organizational factors. The general tendency to favor scientific publication to patent as a mode to encode new knowledge is particularly strong among Italian public researchers, as shown by. Abramo and D'Angelo (2005). Policies and incentive systems adopted by PRIs to foster an entrepreneurial aptitude among their employees have revealed inappropriate or insufficient.

A for the low licensing performance, analysis of the causes is not an easy task, as a number of determinants and, above all, the synergistic effects of their interaction probably concur to determine this situation. Aside from a possible sociological cause, i.e. that Italy productive system is less inclined to taking risks and undergoing changes than US, a low demand by industry is to be excluded. In spite of comparatively low private investments on R&D, a demand for inventions by the Italian industry does exist indeed, as shown by the chronic and growing deficit of Italy technology balance of payments (Tbp), particularly for patents: from €270 million in 1998 to €550 million in 2003. A possible mismatch between public research supply and industrial demand is to be excluded as well. To verify this possibility the correlation between Cnr patents and the Italian Tbp has been measured. First, Cnr filed and licensed patents have been classified according to the Tbp standard classification codes. Then, a correlation analysis between filed and licensed patents with the Tbp in each industrial sector has been carried out. The correlation index between filed patents and Tbp (–0.42), shows a significant correlation between Cnr patent supply and industry demand. In other words, Cnr files more patents in those industrial sectors where the Tbp is more negative. Instead, the correlation index between Cnr patents licensed out and Tbp is –0.30, which shows a weak correlation between industry demand and acquisition of Italian patents. This means that Italian firms tend to satisfy their innovation needs more from abroad than from internal public supply. This finding is confirmed by a recent empirical survey among Italian PRIs by Abramo and D'Angelo (2005). For two thirds of the results with immediate industrial interest (72% of total research output) derived from PRIs high-tech research, there are Italian companies potentially able to exploit them.

While all the conditions seem to be there, technology transfer performance by patent licensing remains extremely poor. In his previous analysis of Cnr licensing activities over a 15-year period, Abramo, 1998, identifies in the technology transfer management policies and practices the main determinant of low performance. No major changes have occurred in the period under exam.

Differently from the US, where there is a long history of university-industry collaboration and technology transfer, Italy has always been characterized by comparatively weak links. Differently from the US, where the Government has enacted laws, such as the Stevenson-Wydler Act and the Bay-Dole Act, to further strengthen such links, in Italy most needed government intervention and consequent institutional building and vision have been quite lacking or insufficient. As a consequence, Italian PRIs policies, organization and management systems, especially incentive systems, are not sufficiently geared towards a strong integration with industry: very few Italian universities have a TLO, and the ones that have it do not usually show adequate capacity (both quantitative and qualitative) and recognition within the organization.

The Italian case shows that public research excellence is not sufficient to assure domestic industry competitiveness and socio-economic development. Making new knowledge available to domestic industry is as important as producing it. While the current debate tend to focus on the scarce resources devoted to R&D, much more attention should be paid on how make those investments more fruitful, by eliminating the public-to-private technology transfer bottleneck.

**Acknowledgements**
The author gratefully acknowledges Daniele Archibugi and Alberto Silvani of Cnr.

Table 1 - Cnr R&D expenditures (€ mln at constant prices 2001); patents filed; and R&D productivity.

| | 1996 | 1997 | 1998 | 1999 | 2000 | 2001 | Total | Average | % |
|---|---|---|---|---|---|---|---|---|---|
| R&D expenditures | 763,93 | 713,65 | 717,75 | 722,02 | 774,72 | 793,00 | 4485,07 | 747,51 | 100% |
| Patents filed | 38 | 30 | 46 | 42 | 39 | 62 | 257 | 42,83 | 100% |
| Source: | | | | | | | | | |
|   Institutional R&D | 30 | 25 | 39 | 34 | 30 | 44 | 202 | 33,67 | 79% |
|   Targeted projects | 8 | 5 | 7 | 8 | 9 | 18 | 55 | 9,17 | 21% |
| Extension: | | | | | | | | | |
|   Italy | 25 | 21 | 33 | 26 | 20 | 46 | 171 | 28,50 | 67% |
|   Pct | 10 | 7 | 13 | 14 | 12 | 11 | 67 | 11,17 | 26% |
|   Epo | 2 | 0 | 0 | 1 | 3 | 0 | 6 | 1,00 | 2% |
|   Uspto | 0 | 2 | 0 | 0 | 1 | 1 | 4 | 0,67 | 2% |
|   Others | 1 | 0 | 0 | 2 | 3 | 3 | 9 | 1,50 | 4% |
| Ownership: | | | | | | | | | |
|   Cnr | 30 | 27 | 33 | 28 | 30 | 50 | 198 | 33,00 | 77% |
|   Cnr and PRIs | 5 | 1 | 5 | 12 | 6 | 10 | 39 | 6,50 | 15% |
|   Cnr and Industry | 3 | 2 | 8 | 2 | 3 | 2 | 20 | 3,33 | 8% |
| R&D productivity | 4,97% | 4,20% | 6,41% | 5,82% | 5,03% | 7,82% | | 5,71% | |

*Table 2 - Cnr patents filed and licensed out disaggregated according to source, extension and ownership - period 1996-2001.*

|  | 1996 | 1997 | 1998 | 1999 | 2000 | 2001 | Total | Average | % |
|---|---|---|---|---|---|---|---|---|---|
| Patents filed | 38 | 30 | 46 | 42 | 39 | 62 | 257 | 42,83 |  |
| Patents licensed out | 12 | 5 | 10 | 5 | 11 | 7 | 50 | 8,33 |  |
| Source: |  |  |  |  |  |  |  |  |  |
|   Institutional R&D | 9 | 4 | 9 | 5 | 9 | 6 | 42 | 7,00 | 84% |
|   Targeted projects | 3 | 1 | 1 | 0 | 2 | 1 | 8 | 1,33 | 16% |
| Extension: |  |  |  |  |  |  |  |  |  |
|   Italy | 6 | 1 | 2 | 2 | 6 | 4 | 21 | 3,50 | 42% |
|   Pct | 4 | 1 | 7 | 3 | 3 | 1 | 19 | 3,17 | 38% |
|   Epo | 0 | 1 | 0 | 0 | 1 | 1 | 3 | 0,50 | 6% |
|   Usa | 0 | 0 | 0 | 0 | 0 | 0 | 0 | 0,00 | 0% |
|   Others | 2 | 2 | 1 | 0 | 1 | 1 | 7 | 1,40 | 14% |
| Ownership: |  |  |  |  |  |  |  |  |  |
|   Cnr | 11 | 4 | 6 | 2 | 10 | 6 | 39 | 6,50 | 78% |
|   Cnr and PRIs | 0 | 0 | 4 | 0 | 0 | 1 | 5 | 0,83 | 10% |
|   Cnr and Industry | 1 | 1 | 0 | 3 | 1 | 0 | 6 | 1,00 | 12% |
| Licensing productivity | 31,58% | 16,67% | 21,74% | 11,90% | 28,21% | 11,29% |  | 19,46% |  |

*Table 3 - Licensing productivity of Cnr patents disaggregated according to source, extension and ownership.*

| Period 1996-2001 | Patents filed | Patents licensed out | Licensing productivity |
|---|---|---|---|
| Patents filed | 257 | 50 | 19,46% |
| Source: | | | |
|    Institutional R&D | 202 | 42 | 20,79% |
|    Targeted projects | 55 | 8 | 14,55% |
| Extension: | | | |
|    Italy | 171 | 21 | 12,28% |
|    Pct | 67 | 19 | 28,36% |
|    Epo | 6 | 3 | 50,00% |
|    Usa | 4 | 0 | 0,00% |
|    Others | 9 | 7 | 77,78% |
| Ownership: | | | |
|    Cnr | 198 | 39 | 19,70% |
|    Cnr and PRIs | 39 | 5 | 12,82% |
|    Cnr and Industry | 20 | 6 | 30,00% |

*Table 4 - Research and licensing productivities comparative analysis Cnr-Mit 1999-2001 (monetary values expressed in million Euro at constant prices 2001). The "%" columns represent the ratio Cnr/Mit.*

|  | 1999 | | | 2000 | | | 2001 | | | Average | | |
|---|---|---|---|---|---|---|---|---|---|---|---|---|
|  | MIT | CNR | % | MIT | CNR | % | MIT | CNR | % | MIT | CNR | % |
| Research expenditure | 695,36 | 721,95 | 103,82% | 746,00 | 775,11 | 103,90% | 764,00 | 793,00 | 103,80% | 735,12 | 763,35 | 103,84% |
| Patented inventions[*] | 143 | 34 | 23,78% | 150 | 34 | 22,67% | 160 | 54 | 33,75% | 151 | 40,67 | 26,93% |
| Research productivity | 20,56% | 4,71% | 22,90% | 20,11% | 4,39% | 21,82% | 20,94% | 6,81% | 32,52% | 20,54% | 5,33% | 25,94% |
| Patent licenses | 87 | 4 | 4,60% | 90 | 10 | 11,11% | 90 | 5 | 5,56% | 89,00 | 6,33 | 7,12% |
| Licensing productivity | 60,84% | 11,76% | 19,34% | 60,00% | 29,41% | 49,02% | 56,25% | 9,26% | 16,46% | 58,94% | 15,57% | 26,42% |
| Quantitative return to R&D investment | 12,51% | 0,55% | 4,43% | 12,06% | 1,29% | 10,69% | 11,78% | 0,63% | 5,35% | 12,11% | 0,83% | 6,85% |
| Revenues | 20,92 | 0,60 | 2,86% | 35,14 | 0,64 | 1,81% | 83,78 | 0,68 | 0,81% | 46,61 | 0,64 | 1,37% |
| Expenditures on patents | 6,20 | 0,42 | 6,81% | 6,31 | 0,33 | 5,28% | 7,24 | 0,41 | 5,66% | 6,58 | 0,39 | 5,90% |
| Gross margin | 14,71 | 0,18 | 1,19% | 28,84 | 0,30 | 1,05% | 76,53 | 0,27 | 0,35% | 40,03 | 0,25 | 0,62% |

*\*Only patents. Trademarks and copyrights are not counted*

*Table 5 - Time-series analysis of Cnr patenting and licensing activities in the periods 1989-1995 and 1996-2001.*

|  | Average 1989-1995 | Average 1996-2001 | Variation |
|---|---|---|---|
| R&D expenditure (million Euro at constant prices 2001) | 824,51 | 747,51 | -9% |
|     Targeted Projects | 130,61 | 106,37 | -19% |
|     Institutional R&D | 693,90 | 641,14 | -8% |
| Patents filed | 73 | 43 | -41% |
|     Patents arisen from Targeted Projects | 34 | 9 | -74% |
|     Patents arisen from Institutional R&D | 39 | 34 | -13% |
| R&D productivity | 8,85% | 5,85% | -34% |
|     Targeted Projects productivity | 26,03% | 8,62% | -67% |
|     Institutional R&D productivity | 5,62% | 5,25% | -7% |
| Patents licensed out | 8,14 | 8,34 | 2% |
| Licensing productivity | 11,15% | 19,40% | 74% |
|     Targeted Projects licensing productivity | 8% | 14,5% | 81% |
|     Institutional R&D licensing productivity | 17,6% | 21% | 19% |
| Patent portfolio | 601 | 557 | -7% |
| Licensing productivity referred to patent portfolio | 1,35% | 1,50% | 11% |
| Quantitative return to R&D investment | 0,99% | 1,12% | 13% |

**References**

Abramo G., 1998, "Il sistema ricerca in Italia: il nodo del trasferimento tecnologico", *Economia e Politica Industriale*, n. 99.

Abramo G., D'Angelo A.C., 2005. La ricerca pubblica in Italia: per chi suona la campana?", Economia Pubblica, XXXV, n. 6.

Abramo G., Pugini F., 2005, *"*L'attività di licensing delle università italiane: un'indagine empirica", forthcoming in *Economia e Politica Industriale.*

Association of University Technology Managers (AUTM), 2004, "Technology licensing survey: Fiscal year 2003".

Balconi M., Breschi S., Lissoni F., 2003, "Il trasferimento di conoscenze tecnologiche dall'università all'industria in Italia: nuova evidenza sui brevetti di paternità dei docenti", in: Bonaccorsi A. (ed.), *Il sistema della ricerca pubblica in Italia*, Franco Angeli.

Baldini N., Grimaldi R., Sobrero M., 2003, "Intervento istituzionale e commercializzazione della conoscenza accademica: uno studio del sistema brevettuale delle università italiane". *XIV Riunione Scientifica della Associazione Italiana di Ingegneria Gestionale su Imprenditorialità e Competenze Manageriali*, Bergamo.

Ben-David, J., 1977, Centers of learning: Britain, France, Germany, and the United States, *McGraw-Hill*, New York.

Bonaccorsi A., Daraio C., 2002, "The organization of science. Size, agglomeration and age effects in scientific productivity", *Conference on Rethinking Science Policy*, 21-23 March, Brighton, UK.

Cohen W.M., Nelson R.R., Walsh J.P., 2002, "Links and Impacts: The influence of public research on industrial R&D", *Management Science*, vol. 48, n. 1.

Colyvas J., Crow M., Geljins A., Mazzoleni R., Nelson R. R., Rosenberg N., Sampat B. N., 2002, "How do university inventions get into practice?", *Management Science*, vol. 48, n. 1.

Commissione Europea, 2002, "Key figures 2002: indicators for benchmarking of national research policies".

Geiger R., 1988, "Research and relevant knowledge: American research universities since World War II", Oxford University Press, New York.

Howells J., McKinlay C., 1999, "Commercialization of university research in Europe", *Report to the Advisory Council on Science and Technology*, Ontario, Canada.

Istat, 2003, "La ricerca e sviluppo in Italia".

Istat, 2003, "L'innovazione nelle imprese italiane negli anni 1998-2000".

Lowe J., 1993, "Commercialization of university research: a policy perspective, *Technology Analysis and Strategic Management*, n. 5(1).

Mowery D.C., Nelson R.R., Sampat B.N., Ziedonis A.A., 2001, "The growth of patenting and licensing by US universities: an assessment of the effects of the Bayh-Dole Act of 1980, *Research Policy*, n. 30.

Mowery D.C., Sampat B. N., 2005, The Bayh-Dole Act of 1980 and university-industry technology transfer: a model for other OECD Governments?, Journal of Technology Transfer, 30.

Nelson, R.R., 2004. The market economy and the scientific commons, Research Policy, 33.

Oecd, 2004, "Main Science and Technology Indicators", vol. 2004/2nd edition.

Orsenigo L., 2002, "Research funds in Italian universities", working paper.

Peterson J., Sharp M., 1998, "Technology policy in the European Union", Macmillan, Basingstoke, U.K.

Piccaluga A., Patrono A., 2001, "Attività brevettuale degli enti pubblici di ricerca italiani. Un'analisi sul periodo 1982-2001", *Economia e Politica Industriale* n.109.

Pressman L., et al., 1995, "Pre-production investment and jobs induced by Mit exclusive patent licenses: A preliminary model to measure the economic impact of university licensing", *Journal of the Association of University Technology Managers*.

Rosenberg N., Nelson R. R., 1994 "American universities and technical advance in industry", *Research Policy* 23.

Siegel D. S., Waldman D., Link A., 2003, "Assessing the impact of organizational practices on the relative productivity of university technology transfer offices: an exploratory study", Research Policy n. 32.

Stokes, D. E. 1977, "Pasteur's quadrant: basic science and technological innovation, Brookings Institution Press, Washington D.C.

Tassey G., 1991, "The Functions of Technology Infrastructure in a Competitive Economy", *Research Policy* 20.

Thursby J.G., Thursby M.C., 2002, "Who is selling the ivory tower? Sources of growth in university licensing", *Management Science*, Vol. 48, n. 1.

Washburn J., 2005. University, Inc.: Corporate Corruption of Higher Education, Basic Books.

[i] The acronym PRIs, public research institutions, refers to both universities and public research laboratories.

[ii] On October 2001 a new law on patents was enacted in Italy, which gave PRIs employees exclusive rights on property and exploitation of their inventions. This new law substituted the previous one, dating back to 1939, which closely resembled the Bayh-Dole Act.

[iii] Why "net"? An invention transferred to industry may displace current technology. Therefore, the assessment of the invention's socio-economic impact should take into account, in addition to the benefits for the innovator and the positive externalities occurring to other firms due to relevant knowledge and market spillovers, also the negative externalities for those firms whose technologies are made obsolete by the innovation.

[iv] A recent re-organisation has aggregated a number of research labs, so the actual number in the period covered in the study was around 330.

[v] Balconi et al. 2002, have shown indeed that the number of patents whose authors belong to academic staff is much higher than those owned by universities. Even accounting for this, Cnr patent productivity results higher than universities.

[vi] Targeted Projects, which last five years, are research programmes jointly carried out by Cnr, other PRIs, and industry on topics reputed as strategic for the country.

[vii] Unless otherwise specified, by patent it is meant all possible forms of intellectual property rights, such as trademarks, software copyrights, etc.

[viii] One single Pct filing allows patent extension to over 100 member countries of the Patent Cooperation Treaty. The same holds true for the European Patent Office filing, as far as the European member countries are concerned.

[ix] At purchasing parity power exchange rate of euro 0,9873 per dollar.

[x] Trademarks and copyrights are not included.

[xi] This figure is conservative, as it regards licenses. Few licences are non exclusive, involving the same patent; others involve more than one patent. The net balance actually shows a number of patents licensed out larger than the number of licenses, as stated by the director of Mit TLO.

[xii] The numerator of the productivity indicator includes, in addition to licenses, also options by companies. Therefore, the result is overestimated.

[xiii] Here too, trademarks and copyrights are not included.

[xiv] For safe of completeness, in the period considered Mit granted 243 software end-use licenses; Cnr only 4.

[xv] That estimate does not include investment and jobs generated by non-exclusive patent agreements, or by no longer active exclusive patent license agreements, or by any type of copyright license agreement.

[xvi] Cnr Executive Committee act n. 305 of May 4, 1995.